\documentclass{svjour2}                    % onecolumn
\smartqed  % flush right qed marks, e.g. at end of proof
\newcommand{\tr}{\mathop{\rm tr}}
\newcommand{\Tr}{\mathop{\rm Tr}}

\newcommand{\ci}{\mathop{\textrm{i}}}

\journalname{General Relativity and Gravitation}
\begin{document}

\title{Obtaining the Weyl tensor from the Bel-Robinson tensor%\thanks{Grants or other notes
}

\titlerunning{Weyl tensor from Bel-Robinson tensor}        % if too long for running head

\author{Joan Josep Ferrando \and \\ Juan Antonio S\'aez }

\authorrunning{J.J. Ferrando \and J.A. S\'aez} % if too long for running head

\offprints{Joan J. Ferrando}          % Insert a name or remove this line
\institute{J.J. Ferrando \at Departament d'Astronomia i
Astrof\'{\i}sica,
\\Universitat
de Val\`encia, E-46100 Burjassot, Val\`encia, Spain.\\
\email{joan.ferrando@uv.es}   \and J.A. S\'aez \at
Departament de Matem\`atiques per a l'Economia i l'Empresa,\\
Universitat de Val\`encia, E-46071 Val\`encia, Spain.\\
\email{juan.a.saez@uv.es}}
\date{Received: date / Revised version: date}

\maketitle

\begin{abstract}

The algebraic study of the Bel-Robinson tensor proposed and
initiated in a previous work (Gen. Relativ. Gravit. {\bf 41}, see
ref [11]) is achieved. The canonical form of the different algebraic
types is obtained in terms of Bel-Robinson eigen-tensors. An
algorithmic determination of the Weyl tensor from the Bel-Robinson
tensor is presented.

\end{abstract}

\section{Introduction}
\label{intro}

The {\it super-energy Bel tensor} is a rank 4 tensor, quadratic in
the Riemann tensor, which plays an analogous role for gravitational
field to that played by the energy tensor for electromagnetism. It
was introduced by Bel to define intrinsic states of gravitational
radiation \cite{bel-1} \cite{bel-2} \cite{bel-3}. In the vacuum
case, the Bel tensor coincides with the {\it super-energy
Bel-Robinson tensor} $T$, built with the same expression by
replacing the Riemann tensor with the Weyl tensor $W$, namely:
\begin{equation}
\label{BR-1} {T_{\alpha  \mu \beta \nu}} = \frac14 \left({{{
W_{\alpha}}^{\rho}}_{\beta}}^{ \sigma} W_{\mu \rho \nu \sigma } +
{{{* W_{\alpha}}^{\rho}}_{\beta}}^{ \sigma} * W_{\mu \rho \nu \sigma
}\right) \, ,
\end{equation}
where $*$ denotes the Hodge dual operator.

In recent years a lot of work has been devoted to analyzing the
properties of the super-energy tensors, and to studying their
generalization to any dimension and to any physical field
\cite{seno}. In a recent paper \cite{garcia}, where the dynamic laws
of super-energy are accurately analyzed, up-to-date references on
the Bel and Bel-Robinson tensors can be found.

Some algebraic properties of the Bel-Robinson tensor (BR tensor)
were studied by Debever early on \cite{debever-1} \cite{debever-2}.
The intrinsic algebraic characterization of a BR tensor was obtained
a few years ago in a paper \cite{bergqvist-lan-1} that presents the
conditions on the BR tensor playing a similar role to that played by
the algebraic Rainich \cite{rai} conditions for the electromagnetic
field.

For a rank 2 tensor satisfying the algebraic Rainich conditions an
exhaustive algebraic study is fully known (see the original work by
Rainich \cite{rai} or the more recent one \cite{fsY}). Nevertheless,
the algebraic study of the BR tensor has not been wholly achieved.
In a recent work \cite{fsBR1} we have posed all the different
aspects of a full algebraic study of the BR tensor:

(I) Algebraic classification of a BR tensor $T$: the symmetries of
the BR tensor $T$ allow us to consider and analyze it as a linear
map on the nine dimensional space of the traceless symmetric
tensors. What algebraic classification follows on from this study?
What relationship exists between these BR classes and the Petrov-Bel
types of the Weyl tensor?

(II) Canonical form of the BR tensor $T$ in terms of its invariant
spaces and scalars: for every algebraic type of the BR tensor, the
eigenvalues and eigenvectors should be analyzed, as well as the
canonical expression of $T$ in terms of them. What relationship
exists between the spaces and the scalar invariants of both the BR
tensor and the Weyl tensor?

(III) Expression of the Weyl tensor in terms of the BR tensor: it is
known that the BR tensor $T$ determines the Weyl tensor $W$ up to a
duality rotation. But the explicit expression of $W$ in terms of $T$
has not been established.

In \cite{fsBR1} we have tackled the algebraic problems of the BR
tensor stated in the three points above: we have solved the first
one and have given preliminary results on points (II) and (III). The
goal of the present work is to fully solve them and to gain a
comprehensive algebraic understanding of the BR tensor.

In section \ref{sec:BR-II} we summarize some results on point (I)
obtained in \cite{fsBR1}: we introduce some notation and present the
classification of the non vanishing BR tensors in nine classes that
can be distinguished considering both, the eigenvalue multiplicity
and the degree of the minimal polynomial. The correspondence of
these classes with the Petrov-Bel types and some type I
degenerations is also outlined.

Section \ref{sec:BR-III} is devoted to studying the eigentensors and
invariant spaces of the BR tensor and to analyze the their
relationship with the Weyl canonical bivectors. The canonical form
of a BR tensor in terms of its eigentensors is also presented.

In section \ref{sec:BR-IV} we study the Segr\`e types of the BR
tensor and offer them in the form of a flow chart that explains the
relationship between the different BR types.

In section \ref{sec:BR-V} we give the algorithms that explicitly
determine (up to a duality rotation) the Weyl tensor in terms of the
BR tensor.

Some notation used in the paper is presented in appendix
\ref{Ap-notation}. In appendix \ref{Ap-frames} we summarize basic
properties of the frames of vectors, bivectors and symmetric
tensors. Appendix \ref{Ap-technical} offers some technical results
that are required in section  \ref{sec:BR-V}.

\section{Algebraic types of the Bel-Robinson tensor}
\label{sec:BR-II}

We shall note $g$ the space-time metric with signature $\{ -, +,+,+
\}$. The conventions of signs are those in the book by Stephani {\em
et al.} \cite{kra}. We take for the Bel-Robinson tensor (BR tensor)
$T$ the expression (\ref{BR-1}), that differs in a factor from the
original one by Bel \cite{bel-3}.

In \cite{fsBR1} we have shown that the algebraic types of the BR
tensor correspond to a refinement of the Petrov-Bel classification.
In order to explain this correspondence, we summarize the algebraic
classification of the Weyl tensor in next subsection.

\subsection{Invariant classification of the Weyl tensor}
\label{subsec:weyl-class}

A self--dual 2--form is a complex 2--form ${\cal F}$ such that
$*{\cal F}= \textrm{i}{\cal F}$. We can associate biunivocally with
every real 2--form $F$ the self-dual 2--form ${\cal
F}=\frac{1}{\sqrt{2}}(F-\textrm{i}*F)$. In short, here we refer to a
self--dual 2--form as a {\it SD bivector}. The endowed metric on the
3-dimensional complex space of the SD bivectors is ${\cal
G}=\frac{1}{2}(G-\textrm{i} \; \eta)$, $\eta$ being the metric
volume element of the space-time, and $G$ being the metric on the
space of 2--forms, $G=\frac{1}{2} g \wedge g$. Here $\wedge$ denotes
the double-forms exterior product (see Appendix
\ref{ap-2tensors}.5).

The algebraic classification of the Weyl tensor $W$ can be obtained
\cite{petrov-W}, \cite{bel-3} by studying the traceless linear map
defined by the self--dual (SD) Weyl tensor ${\cal W}=\frac{1}{2}
(W-\textrm{i}*W)$ on the SD bivectors space. This {\em
SD-endomorphism} (see notation in Appendix \ref{ap-SD}) has
associated the complex scalar invariants $a = \Tr {\cal W}^2$ and $b
= \Tr {\cal W}^3$, and its characteristic equation is of degree
three.

Then, the Petrov-Bel classification follows taking into account both
the eigenvalue multiplicity and the degree of the minimal
polynomial. In the algebraically general case (type I) the
characteristic equation admits three different roots and occurs when
$6\, b^2 \not= a^3$. When $6\, b^2 = a^3 \not=0$ there is a double
root and a simple one, and therefore the minimal polynomial
distinguishes between types D and II. Finally, if $a=b=0$ all the
roots are equal, and so zero, and the Weyl tensor is type O, N or
III, depending on the minimal polynomial.

On the other hand, some classes of algebraically general space-times
have been considered in literature \cite{mcar}. This refinement of
the Petrov-Bel classification is based on the Weyl scalar invariant
$M= \frac{a^3}{b^2} - 6 $. The complex invariant $M$ does not vanish
for type I Weyl tensors. When $M$ is a positive real or infinity,
the Weyl tensor is called of type IM$^+$ or IM$^\infty$,
respectively. These classes have the following geometric
interpretation \cite{mcar} \cite{fsI}: M is a positive real or
infinity if, and only if, the four Debever null directions span a
3-plane. When $M$ is a negative real we have the class IM$^-$, which
has been interpreted in terms of permutability properties of the
null Debever directions \cite{fs-aligned-em}. We denote $I_r$ the
class of non `degenerate' type I Weyl tensors, that is to say, those
with non real invariant $M$.

A detailed analysis of these 'degenerate' algebraically general
classes can be found in \cite{fs-aligned-em}, where they are also
interpreted as the space-times where the electric and magnetic parts
of the Weyl tensor with respect to a (non necessarily time-like)
direction are aligned. These classes also contain the purely
electric and purely magnetic space-times which have been accurately
studied in \cite{fsWem}.

\subsection{Invariant classification of the BR tensor}
\label{subsec:BR-class}

The expression (\ref{BR-1}) of the BR tensor may be written in terms
of the SD Weyl tensor as:
\begin{equation}\label{BR-2}
{T_{\alpha  \mu \beta \nu}} = {{{{\cal
W}_{\alpha}}^{\rho}}_{\beta}}^{ \sigma} \overline{\cal W}_{\mu \rho
\nu \sigma }  \, ,
\end{equation}
where $ \bar{\ } $ stands for complex conjugate.

The BR tensor $T$ is a fully symmetric traceless tensor and,
consequently, it defines a traceless symmetric endomorphism on the
9-dimensional space of the traceless symmetric tensors \cite{fsBR1}.
Expression (\ref{BR-2}) implies that the properties of $T$ as an
endomorphism follow from the properties of ${\cal W}$ as an
endomorphism and, as a consequence, the classes of $T$ are
associated with the classes of ${\cal W}$ presented in subsection
above. A first significant fact is the relationship between the
respective eigenvalues \cite{fsBR1}:
\begin{lemma} \label{lemma-eigenvalues}
The BR tensor $T$ defines an endomorphism of the 9-dimensional space
of the traceless symmetric tensors that admits, generically, three
real eigenvalues $t_i$, and three pairs of complex conjugates
eigenvalues $\tau_i, \overline{\tau}_i$ . In terms of the Weyl
eigenvalues $\rho_i$ we have: $t_i = | \rho_i |^2$, and $\tau_{i} =
\rho_j \overline{\rho}_k$ for $(i,j,k)$ a pair permutation of
(1,2,3).
\end{lemma}

On the other hand, the analysis of the eigenvalue multiplicity leads
to the following result \cite{fsBR1}:
\begin{proposition} \label{propo-eigenvalues}
Taking into account the eigenvalue multiplicity we can distinguish
seven classes of non vanishing BR tensors which are related with the
Weyl tensor types as follows:
\begin{description}
\item[] Type I$_r$ iff $\ T$ has nine different eigenvalues,
three real ones and three pairs of complex conjugate: $\{ t_1, t_2,
t_3, \tau_1, \tau_2, \tau_3, \overline{\tau}_1 , \overline{\tau}_2,
\overline{\tau}_3 \}$.
\item[] Type IM$^-$, $M \neq -6$, iff $\ T$ has six different
eigenvalues, a simple and a double real ones, and two double and two
simple complex conjugate eigenvalues: $\{ t, t, t_3, \tau, \tau,
\tau_3, \overline{\tau}, \overline{\tau}, \overline{\tau}_3 \}$.
\item[] Type IM$^+$ iff $\ T$ has six different real
eigenvalues, three simple ones and three double ones: $\{ t_1, t_2,
t_3, \tau_1, \tau_2, \tau_3, \tau_1 , \tau_2, \tau_3 \}$.
\item[] Type IM$^{[-6]}$ iff $\ T$ has three triple
eigenvalues, one real and a pair of complex conjugate: $\{t , t , t,
\tau, \tau , \tau, \overline{\tau}, \overline{\tau}, \overline{\tau}
\}$.
\item[] Type IM$^{\infty}$ iff $\ T$ has three different real eigenvalues with
multiplicities 2, 2, 5: $\{ t, t, 0, 0 , 0, -t, 0, 0 , -t \}$.
\item[] Types D and II iff $\ T$ has three real eigenvalues with
multiplicities 1,4,4: $\{ 4t, t, t, t, -2t, -2t, t, -2t, -2t \}$.
\item[] Types N and III iff $\ T$ has a sole vanishing eigenvalue with
multiplicity 9: $\{ 0, 0, 0, 0 , 0, 0, 0, 0 , 0 \}$.
\end{description}
\end{proposition}

The above classification of the BR tensor may be refined by
considering the degree of the minimal polynomial \cite{fsBR1}:
\begin{proposition}  \label{propo-minimal}
The Petrov-Bel types of the two last cases in proposition
\ref{propo-eigenvalues} can be distinguished by the degree of the
minimal polynomial of the BR tensor:
\begin{description}
\item[] Types D and II have a minimal polynomial of degree 3 and 6,
respectively.
%\vspace{0.0mm}
%
\item[] Types N and III have a minimal polynomial of degree 2 and 3, respectively.
\end{description}
\end{proposition}

It is worth remarking that the nine types of BR tensor that we find
in the above propositions have a definition in terms of invariant
properties of the BR tensor (eigenvalues and minimal polynomial).
Consequently, these BR types admit an intrinsic characterization
with explicit expressions involving the sole BR tensor. These
expressions and their presentation in an algorithmic form can be
found in \cite{fsBR1}.

\section{Eigentensors and canonical forms of the Bel-Robinson tensor}
\label{sec:BR-III}

The algebraic study of the BR tensor presented in \cite{fsBR1} and
summarized in section above classifies $T$ as an endomorphism.
Nevertheless, a whole study of this endomorphism also requires the
study of its eigenspaces and invariant spaces. Note that, in this
case, an eigenvector of $T$ is a traceless symmetric tensor (in
short, eigentensor).

The analysis of the eigentensors (and other characteristic tensors
in the invariant spaces) of the BR tensor $T$ enables us to
accomplish in this section point (II) stated in the introduction.
Indeed, for every algebraic type, we can obtain the canonical
expression of $T$ in terms of the eigentensors and the eigenvalues.
We also present, for every Bel-Robinson type, the relationship
between the eigentensors of $T$ and the eigenbivectors of the
associated Weyl tensor.

\subsection{Type I}
\label{sec:typeI}

The Weyl tensor is Petrov type I when it admits three different
eigenvalues $\{ \rho_{i} \}$. Then, an orthonormal frame $\{ {\cal
U}_{i} \}$ of SD eigenbivectors exists and, consequently, the Weyl
tensor takes the canonical expression \cite{fms}:
\begin{equation} \label{w-can1}
{\cal W}=- \sum_{i=1}^{3} \rho_{i} \  {\cal U}_{i} \otimes {\cal
U}_{i}
\end{equation}
Thus, in this case we have three {\it canonical bivectors} ${\cal
U}_{i}$, fixed up to sign, and six {\it canonical 2--planes} defined
by the associated 2--forms $U_i$ and $*U_i$.

From (\ref{BR-2}) and (\ref{w-can1}) we can compute the BR tensor.
Then, taking into account lemma \ref{lemma-eigenvalues} and
relations given in Appendix \ref{Ap-frames}, we obtain the
following:
\begin{proposition} \label{prop-can-1}
The BR tensor associated with a type I Weyl tensor admits the
canonical expression
\begin{equation} \label{BR-can1}
T= \sum_{1=1}^3 t_i \, \Pi_i \otimes \Pi_i + \sum_{(ijk)} \tau_i \,
\Pi_{jk} \otimes \Pi_{jk} + \sum_{(ijk)} \overline{\tau}_i \,
\Pi_{kj} \otimes \Pi_{kj}
\end{equation}
where $\Pi_{i} = {\cal U}_i \cdot \overline{\cal U}_i$ are
eigentensors associated with the real eigenvalues $t_i$ and
$\Pi_{jk} = {\cal U}_j \cdot \overline{\cal U}_k\, (=
\overline{\Pi}_{kj})$ are eigentensors associated with the
(generically) complex eigenvalues $\tau_i$, $(i,j,k)$ being a pair
permutation of (1,2,3).
\end{proposition}
Note that $2 \Pi_{i} = U_i^2 + *U_i^2$. Thus $\Pi_{i}$ are the
structure tensors of the three 2+2 almost-product structures defined
by the principal planes.

\subsubsection*{Type I$_r$} \label{subsec:TypeI-regular}

In the most regular case the BR tensor has nine different
eigenvalues (see proposition \ref{propo-eigenvalues}). Thus, we
have:
\begin{proposition} \label{prop-can-1}
A type I$_r$ BR tensor admits the canonical expression {\rm (\ref{BR-can1})}.\\[1mm]
All the eigenvalues are different and $\{\Pi_{i}, \Pi_{jk},
\Pi_{kj}\}$ is the (sole) normalized frame of eigentensors.
\end{proposition}

\subsubsection*{Type IM$^+$} \label{subsec:TypeI(+)}

In this case the BR tensor is like the regular case with $\tau_i =
\overline{\tau}_i$ (see proposition \ref{propo-eigenvalues}). Thus,
we have:
\begin{proposition} \label{prop-can-1+}
A type IM$^+$ BR tensor admits the canonical expression
\begin{equation} \label{BR-can1+}
T=\sum_{1=1}^3 t_i\,  \Pi_i \otimes \Pi_i + \sum_{(ijk)} \tau_i \, (
\Pi_{jk} \otimes \Pi_{jk} + \Pi_{kj} \otimes \Pi_{kj})
\end{equation}
The eigenvalues $t_i$ are simple and are associated with the
eigentensors $\Pi_{i}$, and the (real) eigenvalues $\tau_i$ are
double and are associated with the eigenplanes $S_{i} =
\langle\Pi_{jk}, \Pi_{kj}\rangle$, $(i,j,k)$ being a pair
permutation of (1,2,3).
\end{proposition}

\subsubsection*{Type IM$^{\infty}$} \label{subsec:TypeI(infinit)}

We can see this case as a degeneration of the previous one making
$t_1 = \tau_2 = \tau_3 =0 $ and $ t \equiv t_2 = t_3 = - \tau_1 $
(see proposition \ref{propo-eigenvalues}). Thus, we have:
\begin{proposition} \label{prop-can-1infinit}
A type IM$^{\infty}$ BR tensor admits the canonical expression
\begin{equation} \label{BR-can1-infinit}
T = t \, ( \Pi_2 \otimes \Pi_2 + \Pi_3 \otimes \Pi_3 - \Pi_{23}
\otimes \Pi_{23} - \Pi_{32} \otimes \Pi_{32} )
\end{equation}
The eigenvalues $\pm t$ are double and are associated with the
eigenplanes $S_{+} = \langle\Pi_{2}, \Pi_{3}\rangle$ and $S_{-} =
\langle\Pi_{23}, \Pi_{32}\rangle$, respectively. The
five-dimensional eigenspace $S_{0} = \langle \Pi_{1}, \Pi_{12},
\Pi_{21}, \Pi_{13}, \Pi_{31}\rangle$ is associated with a vanishing
eigenvalue.
\end{proposition}

\subsubsection*{Type IM$^-$} \label{subsec:TypeI(-)}

In this case the BR tensor is like the regular case with $t \equiv
t_1 = t_2$ and $\tau \equiv \tau_1 = \tau_2$ (see proposition
\ref{propo-eigenvalues}). Thus, we have:
\begin{proposition} \label{prop-can-1-}
A type IM$^-$ BR tensor admits the canonical expression
\begin{equation} \label{BR-can1-}
\hspace{-3mm}
\begin{array}{lcl}
T & = & t (\Pi_1 \otimes \Pi_1 + \Pi_2 \otimes \Pi_2) + t_3 \Pi_3
\otimes \Pi_3 + \tau (\Pi_{23} \otimes \Pi_{23}+ \Pi_{31} \otimes
\Pi_{31})\\[1mm] &  + & \overline{\tau} (\Pi_{32} \otimes \Pi_{32} +
\Pi_{13} \otimes \Pi_{13}) + \tau_3 \Pi_{12} \otimes \Pi_{12} +
\overline{\tau}_3 \Pi_{21} \otimes \Pi_{21}
\end{array}
\end{equation}
The eigenvalues $t_3$, $\tau_3$ and $\overline{\tau}_3$ are simple
and are associated with the eigentensors $\Pi_{3}$, $\Pi_{12}$ and
$\Pi_{21}$, respectively. The eigenvalues $t$, $\tau$ and
$\overline{\tau}$ are double and are associated, respectively, with
the eigenplanes $S_{t} = \langle\Pi_{1}, \Pi_{2}\rangle$, $S_{\tau}
= \langle\Pi_{23}, \Pi_{31}\rangle$ and $S_{\overline{\tau}} =
\langle\Pi_{32}, \Pi_{13}\rangle$.
\end{proposition}

\subsubsection*{Type IM$^{[-6]}$} \label{subsec:TypeI([-6])}

We can see this case as a degeneration of the previous one making
$t_3 = t$ and $\tau_3 = \tau$ (see proposition
\ref{propo-eigenvalues}). Thus, we have:
\begin{proposition} \label{prop-can-1-6}
A type IM$^{[-6]}$ BR tensor admits the canonical expression
\begin{equation} \label{BR-can1-6}
\hspace{-3mm}
\begin{array}{lcl}
T & = &  t (\Pi_1 \otimes \Pi_1 +  \Pi_2 \otimes \Pi_2 + \Pi_3
\otimes \Pi_3) + \tau (\Pi_{12}  \otimes
 \Pi_{12} + \Pi_{23}  \otimes \Pi_{23}   + \\[1mm] & + & \Pi_{31}  \otimes
 \Pi_{31}) +  \overline{\tau}  (\Pi_{21}  \otimes
 \Pi_{21}+ \Pi_{32}  \otimes \Pi_{32}  + \Pi_{13}  \otimes
 \Pi_{13})
\end{array}
\end{equation}
Associated with the eigenvalues $t$, $\tau$ and $\overline{\tau}$ we
have, respectively, the three-dimensional eigenspaces $P_{t} =
\langle\Pi_{1}, \Pi_{2}, \Pi_{3}\rangle$, $P_{\tau} =
\langle\Pi_{12}, \Pi_{23}, \Pi_{31}\rangle$ and $P_{\overline{\tau}}
= \langle\Pi_{21}, \Pi_{32}, \Pi_{13}\rangle$.
\end{proposition}

\subsection{Type II}
\label{subsec:TypeII}

A type II Weyl tensor has a simple eigenvalue $-2\, \rho$ and a
double one $\rho$, and it admits the canonical expression
\cite{fms}:
\begin{equation} \label{w-can-II}
{\cal W} =2 \rho \, {\cal U} \otimes {\cal U}- \rho {\cal L}
\stackrel{\sim}{\otimes} {\cal K} + \rho {\cal L} \otimes {\cal L}
\end{equation}
where $\{ {\cal U}, {\cal L} , {\cal K} \}$ is a (Jordan) null frame
of SD bivectors.

The {\em canonical normalized bivector} ${\cal U}$ is the
eigenbivector associated with the simple eigenvalue. Associated with
the double eigenvalue we have the invariant plane $\langle {\cal L}
, {\cal K} \rangle$, the {\em canonical null bivector} ${\cal L}$
being the unique eigendirection in this plane.

As a consequence of lemma \ref{lemma-eigenvalues}, we have $t \equiv
t_2 = t_3 = \tau_1$, $t_1 = 4 t$ and $\tau_2 = \tau_3 = -2t$. Then,
from (\ref{BR-2}) and (\ref{w-can-II}), and taking al into account
relations given in Appendix \ref{Ap-frames}, we can compute the BR
tensor and obtain the following:
\begin{proposition} \label{prop-can-II}
A BR tensor of type II admits the canonical expression
\begin{equation} \label{BR-canII}
\begin{array}{lcl}
T & = &  4 \,t  \, \Pi \otimes \Pi  - 2\, t (
\Omega\stackrel{\sim}{\otimes} {\rm N} + \overline{\Omega}
\stackrel{\sim}{\otimes} \overline{{\rm N}} - {\rm N} \otimes {\rm
N} - \overline{{\rm N}} \stackrel{\sim}{\otimes} \overline{{\rm N}}
) \\[1mm] & + & t (\Lambda \stackrel{\sim}{\otimes} {\rm K} + {\rm M}
\stackrel{\sim}{\otimes} \overline{{\rm M}} - \Lambda
\stackrel{\sim}{\otimes} \overline{{\rm M}} - \Lambda
\stackrel{\sim}{\otimes} {\rm M} + \Lambda \stackrel{\sim}{\otimes}
\Lambda)
\end{array}
\end{equation}
where the (complex) null frame  $\{ \Pi, \Lambda, {\rm K}, {\rm N},
\overline{{\rm N}}, \Omega, \overline{\Omega} , {\rm M},
\overline{{\rm M}} \}$ is defined in terms of the Weyl canonical
frame $\{ {\cal U} , {\cal L}, {\cal K} \}$ as in {\rm
(\ref{stnf})}. The eigenvalue $4t$ is simple and it is associated
with the eigentensor $\Pi$. Associated with the quadruple eigenvalue
$t$ we have the eigentensor ${\rm M}_I$ and the invariant space $Q_t
= \{ \Lambda, {\rm M}_R, {\rm K}\}$, with $\Lambda$ the sole
eigentensor.  Associated with the quadruple eigenvalue $-2t$ we have
the invariant spaces $Q_{R} = \{{\rm N}_R, \Omega_R\}$ with ${\rm
N}_R$ the sole eigentensor, and $Q_{I} = \{{\rm N}_I, \Omega_I\}$
with ${\rm N}_I$ the sole eigentensor.
\end{proposition}
In the above proposition and in that follows, $\Psi_R$ and $\Psi_I$
stand, respectively, for the real and imaginary parts of a complex
symmetric tensor $\Psi$.

\subsection{Type D}
\label{subsec:TypeD}

A type D Weyl tensor also has a simple eigenvalue $-2\, \rho$ and a
double one $\rho$, and it admits the canonical expression
\cite{fms}:
\begin{equation} \label{w-can-D}
{\cal W} =2 \rho \, {\cal U} \otimes {\cal U}- \rho{\cal L}
\stackrel{\sim}{\otimes} {\cal K}
\end{equation}
where $\{ {\cal U}, {\cal L} , {\cal K} \}$ is a null frame of SD
bivectors.

The {\em canonical unitary bivector} ${\cal U}$ is the eigenbivector
associated with the simple eigenvalue. Associated with the double
eigenvalue we have the eigenplane $\langle {\cal L} , {\cal K}
\rangle$ where ${\cal L}$ and ${\cal K}$ define two null
eigendirections.

As a consequence of lemma \ref{lemma-eigenvalues}, we also have $t
\equiv t_2 = t_3 = \tau_1$, $t_1 = 4 t$ and $\tau_2 = \tau_3 = -2t$.
Then, from (\ref{BR-2}) and (\ref{w-can-D}), and taking al into
account relations given in Appendix \ref{Ap-frames}, we can compute
the BR tensor and obtain the following:
\begin{proposition} \label{prop-can-D}
A BR tensor of type D admits the canonical expression
\begin{equation} \label{BR-canII}
T =   4 \,t  \, \Pi \otimes \Pi  - 2\, t (
\Omega\stackrel{\sim}{\otimes} {\rm N} + \overline{\Omega}
\stackrel{\sim}{\otimes} \overline{{\rm N}}) \\[1mm] + t
(\Lambda \stackrel{\sim}{\otimes} {\rm K} + {\rm M}
\stackrel{\sim}{\otimes} \overline{{\rm M}})
\end{equation}
where the (complex) null frame  $\{ \Pi, \Lambda, {\rm K}, {\rm N},
\overline{{\rm N}}, \Omega, \overline{\Omega} , {\rm M},
\overline{{\rm M}} \}$ is defined in terms of the Weyl canonical
frame $\{ {\cal U} , {\cal L}, {\cal K} \}$ as in {\rm
(\ref{stnf})}. The eigenvalue $4t$ is simple and it is associated
with the eigentensor $\Pi$. Associated with the quadruple eigenvalue
$t$ we have the eigenspace $R_t = \{ \Lambda, {\rm K}, {\rm M}_R,
{\rm M}_I\}$.  Associated with the quadruple eigenvalue $-2t$ we
have the eigenspace $R_{-2t} = \{{\rm N}_R, \Omega_R, {\rm N}_I,
\Omega_I\}$.
\end{proposition}

\subsection{Type III}
\label{subsec:TypeIII}

A type III Weyl tensor has a triple vanishing eigenvalue and admits
the canonical expression \cite{fms}:
\begin{equation} \label{w-can-III}
{\cal W} = {\cal U} \stackrel{\sim}{\otimes} {\cal L}
\end{equation}
where the {\em canonical null bivector} ${\cal L}$ is the sole
eigenvector that the Weyl tensor admits. Moreover, a null (Jordan)
frame of SD bivectors $\{{\cal L} , {\cal U},  {\cal K} \}$ may be
completed.

In this case all the BR eigenvalues vanish as a consequence of lemma
\ref{lemma-eigenvalues}. Then, from (\ref{BR-2}) and
(\ref{w-can-III}), and taking into account relations given in
Appendix \ref{Ap-frames}, we can compute the BR tensor and obtain
the following:
\begin{proposition} \label{prop-can-III}
A BR tensor of type III admits the canonical expression
\begin{equation} \label{BR-canIII}
T =   \Lambda \stackrel{\sim}{\otimes} \Pi + {\rm N}
\stackrel{\sim}{\otimes} \overline{{\rm N}}
\end{equation}
where the (complex) null frame $\{ \Pi, \Lambda, {\rm K}, {\rm N},
\overline{{\rm N}}, \Omega, \overline{\Omega} , {\rm M},
\overline{{\rm M}} \}$ is defined in terms of the Weyl canonical
frame $\{ {\cal U} , {\cal L}, {\cal K} \}$ as in {\rm
(\ref{stnf})}. Associated with the sole vanishing eigenvalue we have
a bidimensional eigenspace $E = \langle {\rm M}, \overline{{\rm M}}
\rangle$ and three invariant subspaces: $I_1 = \langle \Pi, \Lambda
, {\rm K} \rangle$, containing the sole eigentensor $\Lambda$; $I_2
= \langle \Omega, \overline{{\rm N}} \rangle$, containing the sole
eigentensor $\overline{{\rm N}}$; and $I_3 = \langle
\overline{\Omega} , {{\rm N}} \rangle $, containing the sole
eigentensor ${\rm N}$.
\end{proposition}

\subsection{Type N}
\label{subsec:TypeN}

A type N Weyl tensor also has a triple vanishing eigenvalue and it
admits the canonical expression \cite{fms}:
\begin{equation}  \label{w-can-N}
{\cal W} = {\cal L} \otimes {\cal L}
\end{equation}
Now, the SD bivectors orthogonal to the {\em canonical null
bivector} ${\cal L}$ define a two-dimensional eigenspace.

In this case all the BR eigenvalues also vanish as a consequence of
lemma \ref{lemma-eigenvalues}. Then, from (\ref{BR-2}) and
(\ref{w-can-III}), and taking al into account relations given in
Appendix \ref{Ap-frames}, we can compute the BR tensor and obtain
the following:
\begin{proposition}  \label{prop-can-N}
A BR tensor of type N admits the canonical expression
\begin{equation} \label{BR-canN}
T= \Lambda \otimes \Lambda
\end{equation}
where $\Lambda = {\cal L} \cdot \overline{{\cal L}}$, ${\cal L}$
being the canonical null bivector. Associated with the sole
vanishing eigenvalue we have the eight-dimensional eigenspace
orthogonal to $\Lambda$.
\end{proposition}

\section{Segr\`e types of the Bel-Robinson tensor}
\label{sec:BR-IV}

Propositions \ref{propo-eigenvalues} and \ref{propo-minimal}
summarize the eigenvalue multiplicity and the degree of the minimal
polynomial of the different BR classes. Although these results
restrict the Segr\`e type, they do not fix them for the BR classes
II, III and N. The study of the eigentensors given in section above
completes all the information on these Segr\`e types.

We present the Segr\`e types of the BR classes as a flow chart that
helps us to visualize the different kinds of degeneration. The
continuous arrows correspond to a degeneration in the eigenvalue
multiplicity, the four levels having, respectively, nine (type
I$_r$), six (types IM$^+$ and IM$^-$), three (types I$^{\infty}$,
IM$^{[-6]}$, II and D) or one (types III, N and O) different
eigenvalues. The dash arrows correspond to a degeneration in the
degree of the minimal polynomial: nine (type I$_r$), six (types
IM$^+$, IM$^-$ and II), three (types I$^{\infty}$, IM$^{[-6]}$, D
and III), two (type N) or one (type O).

\newpage

\vspace*{2.5cm}
%
%\hspace*{0mm}
%
{\scriptsize
 \setlength{\unitlength}{0.7cm}
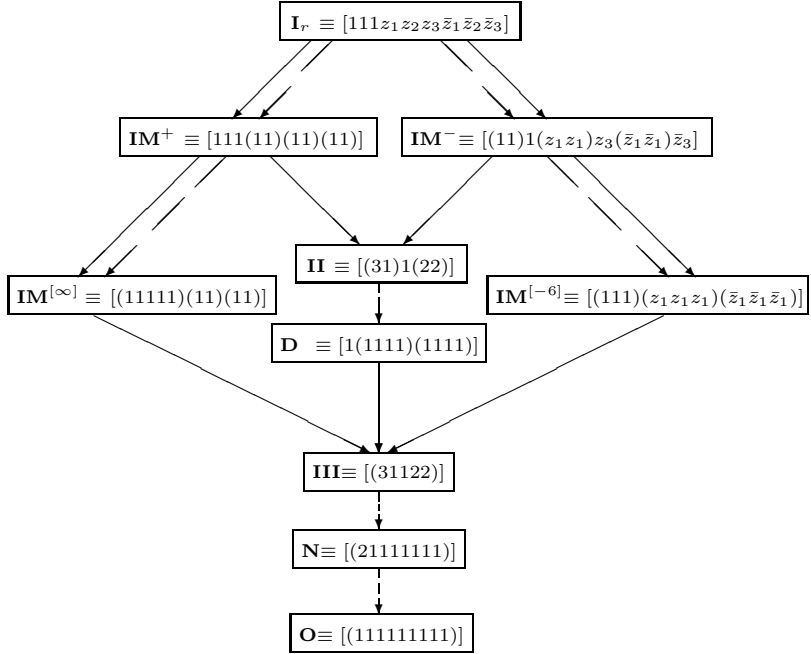
\noindent \hspace*{1.7mm} \vspace*{2.5cm}
\begin{picture}(10,14)
%\thicklines
 \put(5,17){\framebox(4.5,0.7)}
 \put(5.2,17.2) {\bf{I$_r $} {{$\equiv[ 111 z_1 z_2 z_3
\bar{z}_1 \bar{z}_2 \bar{z}_3 ]$}}}

\put(7.3,14.8){\framebox(5.8,0.7)}
\put(7.45,15){\bf{IM$^-$}{{$\equiv[ (11)1 (z_1 z_1) z_3 (\bar{z}_1
\bar{z}_1) \bar{z}_3 ]$}}}

\put(2,14.8){\framebox(4.8,0.7)} \put(2.15,15){\bf{IM$^+ $}
{{$\equiv[ 111 (11) (11) (11) ]$}}}

\put(8.9,11.8){\framebox(6,0.7)}
 \put(9.05,12){\bf{IM$^{[-6]}
$}{{$\equiv[  (111) (z_1 z_1 z_1) (\bar{z}_1 \bar{z}_1 \bar{z}_1)
]$}}}

\put(5.3,12.4){\framebox(3.2,0.7)} \put(5.5,12.6){\bf{II}{ $\equiv[
(31)1 (22) ]$}}

\put(-0.1,11.8){\framebox(5,0.7)} \put(0.05,12){\bf{IM$^{[\infty]}
\equiv$}{{ $[(11111) (11) (11) ]$}}}

\put(5,11.1){\bf{D} { $\equiv [ 1(1111)(1111) ]$}}
\put(4.85,10.9){\framebox(4,0.7)}

\put(5.6,8.7){\bf{III}{$\equiv[ (31122) ]$}}
\put(5.45,8.45){\framebox(2.8,0.7)}

\put(5.4,7.2){\bf{N}{$\equiv[ (21111111) ]$}}
\put(5.25,7){\framebox(3.2,0.7)}

\put(5.2,5.4){\framebox(3.4,0.7)} \put(5.35,5.6){{\bf{O}{$\equiv[
(111111111) ]$}}}

\put(1.5,11.77){\vector(2,-1){5.2}}

\put(12.2,11.77){\vector(-2,-1){5.2}}

\put(3.5,14.8){\vector(-1,-1){2.3}}

\put(4,14.8){\line(-1,-1){0.6}} \put(3.2,14){\line(-1,-1){0.6}}
%\put(2.4,13.2){\line(-1,-1){0.6}}
\put(2.3,13.1){\vector(-1,-1){0.6}}

\put(10.5,14.8){\vector(1,-1){2.3}}

\put(10,14.8){\line(1,-1){0.6}} \put(10.8,14.){\line(1,-1){0.6}}
%\put(11.2,13.6){\line(1,-1){0.5}}
\put(11.7,13.1){\vector(1,-1){0.6}}

\put(8.5,17){\vector(1,-1){1.5}}

\put(8,17){\line(1,-1){0.6}} \put(8.8,16.1){\vector(1,-1){0.6}}
\put(5.6,17){\vector(-1,-1){1.5}} \put(6,17){\line(-1,-1){0.6}}
\put(5.2,16.1){\vector(-1,-1){0.6}}

\put(4.8,14.8){\vector(1,-1){1.7}}
\put(9.,14.8){\vector(-1,-1){1.7}}

%\put(5,16.8){\vector(-1,-1){1.4}} \put(8.5,16.8){\vector(1,-1){1.4}}

%\put(5.5,16.8){\line(-1,-1){.4}} \put(5,16.3){\line(-1,-1){0.4}}
%\put(4.5,15.8){\vector(-1,-1){4.4}}

%\put(8.1,16.2){\line(1,-1){0.44}}
%\put(8.6,15.7){\vector(1,-1){0.2}}

\put(6.85,8){\vector(0,-1){0.3}} \put(6.85,8.18){\line(0,-1){0.1}}
\put(6.85,8.45){\line(0,-1){0.2}}

\put(6.85,6.4){\vector(0,-1){0.3}} \put(6.85,6.7){\line(0,-1){0.2}}
\put(6.85,7){\line(0,-1){0.2}} \put(6.85,10.9){\vector(0,-1){1.75}}
\put(6.85,11.9){\vector(0,-1){0.3}}
\put(6.85,12.4){\line(0,-1){0.2}} \put(6.85,12.1){\line(0,-1){0.2}}

\put(6.85,10.9){\vector(0,-1){1.75}}
\end{picture}}

\vspace{-6cm}

\section{The Weyl tensor in terms of the Bel-Robinson tensor}
\label{sec:BR-V}

The BR tensor can be obtained from the Weyl tensor by means of the
quadratic expression (\ref{BR-1}). This expression is known to be
invariant under duality rotation of the Weyl tensor. Thus, we can
pose the following question: can the Weyl tensor be determined, up
to a duality rotation, from the BR tensor? or, to be more precise,
is there an explicit algorithm to obtain it?

In \cite{fsBR1} we have solved this problem for the algebraic types
where neither $a$ nor $b$ vanish. In this 'generic' case we give the
Weyl tensor as an explicit concomitant of the BR tensor.

Here we deal with the cases when $a$ or $b$ vanish. This condition
leads to the algebraic types N, III, IM$^{\infty}$ and IM$^{-6}$,
which we study separately. In these 'non generic' cases our approach
is based on three steps. First, we express the BR eigentensors in
terms of the BR tensor; secondly we obtain the Weyl canonical
bivectors from the BR eigentensors; and, finally, we use the Weyl
canonical form to get the Weyl tensor.

In the second step, we will use the following three lemmas that may
be easily shown from the expressions given in Appendix
\ref{Ap-frames}:

\begin{lemma} \label{lemma-L}
Let us consider the symmetric tensor $\Lambda = {\cal L} \cdot
\overline{{\cal L}}$, where ${\cal L}$ is a null SD bivector. Then,
$\Lambda$ determines ${\cal L}$ up to a phase $\phi$ as
\begin{equation} \label{L-de-Lambda}
\begin{array}{l}
{\cal L} = e^{\ci \phi} {\cal L}_0 \, , \quad {\cal L}_0 \equiv
\frac{1}{\sqrt{2}} (L - \ci *L) \, ,
\quad L \equiv l \wedge p \, ,\\[2mm]
l \equiv - \frac{\Lambda(u)}{\sqrt{\Lambda(u,u)}} \, , \quad p
\equiv \frac{x}{\sqrt{x^2}} \, , \quad x \equiv *(l \wedge u \wedge
w) \, ,
\end{array}
\end{equation}
where $u$ is an arbitrary time-like vector and $w$ an arbitrary
vector such that $x \neq 0$.
\end{lemma}
\begin{lemma} \label{lemma-L-U}
Let us consider the symmetric tensors $\Lambda = {\cal L} \cdot
 \overline{{\cal L}}$ and $\Pi = {\cal U} \cdot \overline{{\cal U}}$, where
${\cal L}$ and ${\cal U}$ are two orthogonal SD bivectors, null and
unitary, respectively. Then, $\Lambda$ and $\Pi$ determine ${\cal
L}$ (up to a phase $\phi$) and ${\cal U}$ as
\begin{equation} \label{LU-de-LambdaPi-a}
\begin{array}{l}
{\cal L} = e^{\ci \phi} {\cal L}_0 \, , \quad {\cal L}_0 \equiv
\frac{1}{\sqrt{2}} (L - \ci *L) \, , \quad L \equiv l \wedge p \, ,\\[2mm]
l \equiv - \frac{\Lambda(u)}{\sqrt{\Lambda(u,u)}} \, , \quad p
\equiv \frac{h(w)}{\sqrt{h(w,w)}} \, , \quad h \equiv \frac12 g -
\Pi \, ,
\end{array}
\end{equation}
\begin{equation} \label{LU-de-LambdaPi-b}
%\begin{array}{l}
{\cal U} =  \frac{1}{\sqrt{2}} (U - \ci *U) \, , \quad U \equiv l
\wedge e \, , \quad e \equiv -\frac{v(u)}{v(l,u)} \, , \quad v
\equiv \frac12 g + \Pi  \, ,
%\end{array}
\end{equation}
where $u$ is an arbitrary time-like vector and $w$ an arbitrary
vector such that $h(w) \neq 0$.
\end{lemma}
\begin{lemma} \label{lemma-UU}
Let us consider the symmetric tensor  $\Pi = {\cal U} \cdot
\overline{{\cal U}}$, where ${\cal U}$ is a unitary SD bivector.
Then, $\Pi$ determine the tensorial square of ${\cal U}$ as
\begin{equation} \label{U-de-Pi}
\begin{array}{l}
{\cal U} \otimes {\cal U} = \frac{1}{2} (D - \ci *D) \, , \quad D
\equiv - [ \Pi \wedge \Pi + \frac14 \, g \wedge g]  \, .
\end{array}
\end{equation}
\end{lemma}

We summarize the results of the 'generic' case in the next
subsection. Afterwards, in the following subsections, we study the
four non generic cases separately.

\subsection{Generic case}
\label{subsec:V-general}

The condition $a \not=0 \not= b$ can be stated in terms of the BR
tensor as $\Tr T^2 \not= 0 \not= \tr T^3$. In \cite{fsBR1} we have
shown the following
\begin{theorem} \label{theore-regular}
If $T$ is a BR tensor and
   $\Tr T^2 \neq 0$, $\Tr T^3 \neq 0$, then the
  Weyl tensor can be obtained, up to duality rotation, as
\begin{equation} \label{W(T)}
{\cal W} = e^{\ci \theta}{\cal W}_0 \, , \qquad {\cal W}_0 =
\frac{1}{\sqrt{\Tr T^2 \tr T^3}} \left[4 \, {\cal T}_{(3)} +
\frac{1}{3}\, \Tr T^3 \,{\cal G}\right]
\end{equation}
\begin{equation}
{\cal T}_{(3) \alpha \beta \rho \sigma} \equiv  (T^3)_{\alpha \mu
\nu \beta} \, {\cal G}^{\mu \nu}_{\ \ \, \rho \sigma}
\end{equation}
\end{theorem}

\subsection{Type N}
\label{subsec:V-N}

A BR tensor of type N is characterized by the condition $T^2 = 0$
\cite{fsBR1}. As stated in subsection \ref{subsec:TypeN}, in this
case the Weyl tensor and the BR tensor take, respectively, the
expressions (\ref{w-can-N}) and (\ref{BR-canN}), where $\Lambda =
{\cal L} \cdot  \overline{{\cal L}}$. Then, in a first step, we must
obtain $\Lambda$ in terms of the BR tensor $T$. A direct calculation
leads to the following result.
\begin{lemma} \label{lemma-N}
If  $T = \Lambda \otimes \Lambda$, then
\begin{equation} \label{lambda-N}
\Lambda = \frac{T(X)}{\sqrt{T(X,X)}} \, ,
\end{equation}
where $X=u \otimes u$, $u$ being an arbitrary time-like vector.
\end{lemma}

Secondly, we can obtain ${\cal L}$ (up to a phase $\phi$) in terms
of $\Lambda$ by using lemma \ref{lemma-L}. The effect of this
freedom is a duality rotation on the Weyl tensor obtained as
(\ref{w-can-N}). Then, we have:
\begin{theorem} \label{theorem-N}
If $T$is a BR tensor such that $T^2 = 0$, then the Weyl tensor can
be obtained, up to duality rotation, as
\begin{equation} \label{W(T)}
{\cal W} = e^{\ci \theta}{\cal W}_0 \, , \qquad {\cal W}_0 = {\cal
L}_0 \otimes {\cal L}_0 \, ,
\end{equation}
where ${\cal L}_0$ can be obtained as {\rm (\ref{L-de-Lambda})} in
terms of the BR concomitant $\Lambda$ given in {\rm
(\ref{lambda-N})}.
\end{theorem}

\subsection{Type III}
\label{subsec:V-III}

A BR tensor of type III is characterized by the conditions $T^3 =
0$, $T^2 \not=0$ \cite{fsBR1}. As stated in subsection
\ref{subsec:TypeIII}, in this case the Weyl tensor and the BR tensor
take, respectively, the expressions (\ref{w-can-III}) and
(\ref{BR-canIII}), where $\Lambda = {\cal L} \cdot  \overline{{\cal
L}}$ and $\Pi = {\cal U} \cdot \overline{{\cal U}}$. Then, in a
first step, we must obtain $\Lambda$ and $\Pi$ in terms of the BR
tensor $T$.

From (\ref{BR-canIII}) we can compute the square $T^2$ and obtain:
\begin{equation} \label{sqBRIII}
T^2 = \Lambda \otimes \Lambda
\end{equation}
On the other hand, taking into account expressions given in Appendix
\ref{Ap-frames}, we can also compute the tensor
\begin{equation} \label{BRIII-S-a}
S_{\alpha \beta \mu \nu \rho \sigma} \equiv (T \cdot T)_{\alpha
\beta \rho \sigma \mu \nu} + (T \cdot T)_{\alpha \beta \sigma \rho
\mu \nu} \, ,
\end{equation}
and we obtain:
\begin{equation} \label{BRIII-S-b}
S = ({\rm N} \stackrel{\sim}{\otimes} \bar{{\rm N}} )\otimes \Lambda
+ (\Lambda \stackrel{\sim}{\otimes} \Pi )\otimes \Lambda +
\frac{1}{2} \Lambda \otimes \Lambda \otimes g
\end{equation}
From (\ref{BR-canIII}), (\ref{sqBRIII}) and (\ref{BRIII-S-b}) we
have:
\begin{equation} \label{BRIII-R-a}
S_{\alpha \beta \mu \nu \rho \sigma}- T_{\alpha \beta \mu \nu}
\Lambda_{\rho \sigma} - \frac{1}{2} (T^2)_{\alpha \beta \mu \nu}
g_{\rho \sigma} = 2 R_{\alpha \beta \mu \nu \rho \sigma}
\end{equation}
where $R$ takes the expression:
\begin{equation} \label{BRIII-R-b}
R_{\alpha \beta \mu \nu \rho \sigma} = \Pi_{\alpha \beta }
(T^2)_{\mu \nu \rho \sigma} + (T^2)_{\alpha \beta  \rho \sigma}
\Pi_{\mu \nu}
\end{equation}

Now, from (\ref{sqBRIII}) and (\ref{BRIII-R-b}) we can obtain
$\Lambda$ and $\Pi$ in terms of the BR concomitants $T^2$ and $R$
and we arrive to the following:
\begin{lemma} \label{lemma-III}
If  $T = \Lambda \stackrel{\sim}{\otimes} \Pi + {\rm N}
\stackrel{\sim}{\otimes} \bar{{\rm N}}$, then
\begin{equation} \label{lambda-Pi-III}
\displaystyle
\begin{array}{l}
\displaystyle \Lambda = \frac{T^2 (X)}{\sqrt{T^2(X,X)}} \, , \\[3mm]
\displaystyle \Pi=\frac{1}{T^2(X,X)} \left[ R(X,X)-
\frac{R(X,X,X)}{2 \, T^2(X,X)} T^2(X) \right] \, ,
\end{array}
\end{equation}
where $X=u \otimes u$, $u$ being an arbitrary time-like vector, and
$R$ is given by:
\begin{equation} \label{lambda-Pi-III-b}
\hspace{-0.15cm} \displaystyle 2  R_{\alpha \beta \mu \nu \rho
\sigma} \equiv (T \cdot T)_{\alpha \beta \rho \sigma \mu \nu} + (T
\cdot T)_{\alpha \beta \sigma \rho \mu \nu} - T_{\alpha \beta \mu
\nu} \Lambda_{\rho \sigma} - \frac{1}{2} (T^2)_{\alpha \beta \mu
\nu} g_{\rho \sigma} \, .
\end{equation}
\end{lemma}

In a second step, we can obtain ${\cal L}$ (up to a phase $\phi$)
and ${\cal U}$ in terms of $\Lambda$ and $\Pi$ by using lemma
\ref{lemma-L-U}. The effect of the freedom $\phi$ is a duality
rotation on the Weyl tensor obtained as (\ref{w-can-III}). Then, we
have:
\begin{theorem} \label{theorem-III}
If $T$is a BR tensor such that $T^3 = 0$ and $T^2 \not=0$, then the
Weyl tensor can be obtained, up to duality rotation, as
\begin{equation} \label{W(T)}
{\cal W} = e^{\ci \theta}{\cal W}_0 \, , \qquad {\cal W}_0 =  {\cal
U} \stackrel{\sim}{\otimes} {\cal L}_0  \, ,
\end{equation}
where ${\cal L}_0$ and ${\cal U}$ can be obtained as {\rm
(\ref{LU-de-LambdaPi-a})} and  {\rm (\ref{LU-de-LambdaPi-b})} in
terms of the BR concomitants $\Lambda$ and $\Pi$ given in {\rm
(\ref{lambda-Pi-III}-\ref{lambda-Pi-III-b})}.
\end{theorem}

\subsection{Type IM$^{\infty}$}
\label{subsec:V-IM-infinit}

A BR tensor of type IM$^{\infty}$ is characterized by the conditions
$\Tr T^3 = 0$, $\Tr T^2 \not=0$ \cite{fsBR1}. Then, the Weyl tensor
has a vanishing eigenvalue and the canonical form (\ref{w-can1})
becomes:
\begin{equation} \label{w-can-I-infinit}
{\cal W} = \rho  \left( {\cal U}_2 \otimes {\cal U}_2 - {\cal U}_3
\otimes {\cal U}_3 \right) \, .
\end{equation}
On the other hand, as stated in proposition \ref{prop-can-1infinit},
the BR tensor takes the expression (\ref{BR-can1-infinit}), where
$\Pi_{i} = {\cal U}_i \cdot \overline{\cal U}_i$ and $\Pi_{jk} =
{\cal U}_j \cdot \overline{\cal U}_k\, (= \overline{\Pi}_{kj}), \ j
\not=k$.

In a first step, we must obtain some concomitants of the
eigentensors $\Pi_i$ in terms of the BR tensor $T$. From
(\ref{BR-can1-infinit}), we can compute $T^2$ and then obtain
\begin{equation}\label{bel1a3}
P \equiv \frac{1}{t}\,  T + \frac{2}{t^2}\, T^2 = \Pi_2 \otimes
\Pi_2 + \Pi_3 \otimes \Pi_3
\end{equation}
Using expressions given in Appendix \ref{Ap-frames} we can compute
the tensor:
\begin{equation} \label{bel1a4}
V_{\alpha \beta \mu \nu} \equiv 2 \, P_{\alpha \rho \mu \sigma}
P_{\beta \ \nu}^{\ \rho \ \sigma} - \frac{1}{4} g_{\alpha \beta}
g_{\mu \nu}
\end{equation}
and we arrive to the projector on the eigendirection $\Pi_1$:
\begin{equation} \label{bel1a4b}
V =  \Pi_1 \otimes \Pi_1 \, .
\end{equation}
Then, we can compute:
\begin{equation} \label{bel1a4c}
Q \equiv \Pi_1 \cdot P = \Pi_3 \otimes \Pi_2 +  \Pi_2 \otimes \Pi_3
\end{equation}
Now, from (\ref{bel1a3}) and (\ref{bel1a4c}) we can apply Lemma \ref{lemma-Ac1}
in Appendix C.1 and arrive to the following:
\begin{lemma} \label{lemma-I-infinit}
For a BR tensor $T$ of type IM$^{\infty}$, let $\{\Pi_i\}$ be its
three first normalized eigentensors. Then we have:
\begin{equation} \label{Pi-I-infinit-a}
\displaystyle \Pi_1 = \frac{V (X)}{\sqrt{V(X,X)}} \, ,
\end{equation}
\begin{equation}
\displaystyle \Pi_2 \otimes \Pi_2 -  \Pi_3 \otimes \Pi_3 =
\frac{P(X) \otimes P(X) - Q(X) \otimes Q(X)}{\sqrt{[P(X,X)]^2 -
[Q(X,X)]^2}} \, ,   \label{Pi-I-infinit-a2}
\end{equation}
where $X=u \otimes u$, $u$ being an arbitrary time-like vector such
that $\Pi_1(u) \not= u$, and
\begin{equation}  \label{Pi-I-infinit-b}
P \equiv \frac{1}{t}\,  T + \frac{2}{t^2}\, T^2  \, , \quad Q \equiv
\Pi_1 \cdot P \, , \quad  V_{\alpha \beta \mu \nu} \equiv 2 \,
P_{\alpha \rho \mu \sigma} P_{\beta \ \nu}^{\ \rho \ \sigma} -
\frac{1}{4} g_{\alpha \beta} g_{\mu \nu} \, .
\end{equation}
\end{lemma}

On the other hand, by using lemma \ref{lemma-UU} we obtain:
\begin{equation} \label{Pi-I-infinit-c}
{\cal U}_2 \otimes {\cal U}_2 - {\cal U}_3 \otimes {\cal U}_3 =
\frac{1}{2} (D - \ci *D) \, , \qquad D \equiv \Pi_3 \wedge \Pi_3 -
\Pi_2 \wedge \Pi_2 \, .
\end{equation}
Moreover, from lemma \ref{lemma-eigenvalues} we have $\rho =  e^{\ci
\theta} \sqrt{t}$, where $\theta$ is an arbitrary phase that gives a
duality rotation in the expression (\ref{w-can-I-infinit}) of the
Weyl tensor. Then, taking into account lemma \ref{lemma-I-infinit}
and expression (\ref{Pi-I-infinit-c}) we have:
\begin{theorem} \label{theorem-I-infinit}
If $T$is a BR tensor such that $\Tr T^3 = 0$ and $4 t^2 \equiv \Tr
T^2 \not=0$, then the Weyl tensor can be obtained, up to duality
rotation, as
\begin{equation} \label{W(T)}
W = \cos \theta \, W_0 +  \sin \theta \, *W_0 \, , \quad W_0 =
\sqrt{t} \, \frac{P(X) \wedge P(X) - Q(X) \wedge
Q(X)}{\sqrt{[P(X,X)]^2 - [Q(X,X)]^2}} ,
\end{equation}
where $X=u \otimes u$, $u$ being an arbitrary time-like vector such
that $\Pi_1(u) \not= u$, and where $\Pi_1$ is given in {\rm
(\ref{Pi-I-infinit-a})} and $P$, $Q$ and $V$ are given in {\rm
(\ref{Pi-I-infinit-b})}.
\end{theorem}

\subsection{Type IM$^{[-6]}$}
\label{subsec:V-IM-[-6]}

A BR tensor of type IM$^{[-6]}$ is characterized by the conditions
$\Tr T^3 \not= 0$, $\Tr T^2 =0$ \cite{fsBR1}. Then, the three Weyl tensor
eigenvalues differ in a cubic root of the unity, and the canonical form (\ref{w-can1})
becomes:
\begin{equation} \label{w-can-I-6}
{\cal W} = \rho  \left( \lambda_1 \, {\cal U}_1 \otimes {\cal U}_1 +
\lambda_2 \, {\cal U}_2 \otimes {\cal U}_2 + \lambda_3 \, {\cal U}_3
\otimes {\cal U}_3 \right) \, ,
\end{equation}
where $\lambda_i^3 = 1$. On the other hand, as stated in proposition
\ref{prop-can-1-6}, the BR tensor takes the expression
(\ref{BR-can1-6}). This expression may be written:
\begin{equation}
\label{tipoia} T = t( A + \overline{\lambda}_1 \lambda_2 \,  L +
\lambda_1 \overline{\lambda}_2 \, \bar{L}) \, ,
\end{equation}
where $\, L=\Pi_{13} \otimes \Pi_{13} + \Pi_{32} \otimes \Pi_{32}+
\Pi_{21} \otimes\Pi_{21}$ and
\begin{equation}
\label{tipoia}
A = \Pi_1 \otimes \Pi_1 + \Pi_2 \otimes \Pi_2+ \Pi_3 \otimes
\Pi_3 \, .
\end{equation}

In a first step, we must obtain some concomitants of the
eigentensors $\Pi_i$ in terms of the BR tensor $T$. From
(\ref{tipoia}), and taking into account expressions given in
Appendix \ref{Ap-frames}, we can compute the powers of $T$:
$$T^2 = t^2(A + \lambda_1 \overline{\lambda}_2  \, \ L +  \overline{\lambda}_1 \lambda_2 \,
\bar{L}) \, , \qquad T^3 = t^3(A + L + \bar{L})$$
From these expressions and (\ref{tipoia}) we obtain:
\begin{equation} \label{AdeT}
A = \frac13 \left[\frac{1}{t} T + \frac{1}{t^2}  \, T^2 +
\frac{1}{t^3} \, T^3\right] \, .
\end{equation}
Moreover, we can define the quadratic and cubic concomitants of $A$
given by:
\begin{equation}\label{acero1}
B_{\alpha \beta \gamma \delta \lambda \mu} = 2 \, {A_{\alpha \beta
\lambda}}^{\epsilon} \ A_{\gamma \delta \mu \epsilon} - \frac{1}{2}
A_{\alpha \beta \gamma \delta} g_{\lambda \mu} \, ,
\end{equation}
\begin{equation}\label{acero1b}
C_{\alpha \beta \gamma \delta \lambda \mu \nu \rho} = 2 \,
{A_{\alpha \beta \nu}}^{\epsilon} \, B_{\gamma \delta \lambda \mu
\rho  \epsilon} - \frac{1}{2} \, B_{\alpha \beta \gamma \delta
\lambda \mu} g_{\nu \rho} \, ,
\end{equation}
and, if we compute them taking into account expressions given in
Appendix \ref{Ap-frames}, we obtain:
\begin{equation}\label{aceroBC}
B= \sum_{\sigma \in \Sigma_3} \Pi_{\sigma(1)} \otimes
\Pi_{\sigma(2)} \otimes \Pi_{\sigma(3)} \, , \quad C=\sum_{i<j} [
\Pi_i \stackrel{\sim}{\otimes} \Pi_j ] \otimes [ \Pi_i
\stackrel{\sim}{\otimes} \Pi_j] \, .
\end{equation}
Now, from (\ref{tipoia}) and (\ref{aceroBC}) we can apply Lemma \ref{lemma-Ac2}
in Appendix C.2 and arrive to the following:
\begin{lemma} \label{lemma-I-6}
For a BR tensor $T$ of type IM$^{[-6]}$, let $\{\Pi_i\}$ be its
three first normalized eigentensors. Then the tensor $H= \lambda_1
\, \Pi_1 \otimes \Pi_1 + \lambda_2 \, \Pi_2 \otimes \Pi_2 +
\lambda_3 \, \Pi_3 \otimes \Pi_3$, where $\lambda_i$ are the three
cubic roots of the unity, can be obtained in terms of the BR
concomitants $A$, $B$ and $C$ given in {\rm (\ref{AdeT})}, {\rm
(\ref{acero1})} and {\rm (\ref{acero1b})} as:
\begin{equation} \label{H}
H = \frac{1}{2w \Delta}\left[-(\alpha w^3 +2 r \alpha^2 +r^2)A
+(w^3+r\alpha)E+ r I \right] \,
\end{equation}
where
\begin{equation}
E \equiv A(X) \otimes A(X) - \frac{1}{2} C(X,X) , \quad I \equiv
\frac{1}{4} B(X,X) \otimes B(X,X) - \beta \,  B(X) \, ,
\end{equation}
\begin{equation}
\alpha \equiv \frac13 A(X,X),  \quad \beta \equiv \frac{1}{6}
B(X,X,X) \ \quad \gamma \equiv \frac{1}{12} C(X,X,X,X) \, ,
\end{equation}
\begin{equation}
\Delta \equiv \sqrt{s^2+r^3} \, , \quad r \equiv \gamma-\alpha^2\, ,
\quad s \equiv \frac12(3 \alpha \gamma- \beta^2)- \alpha^3 \, ,
\end{equation}
and where $w$ is a nonvanishing scalar defined by one of the
expressions $w^3 \equiv - s \pm \Delta$, and $X$ being an arbitrary
symmetric tensor such that $\Delta \not= 0$.
\end{lemma}

On the other hand, by using lemma \ref{lemma-UU} we obtain:
\begin{equation} \label{Pi-I-6-a}
\begin{array}{l}
\lambda_1 \, {\cal U}_1 \otimes {\cal U}_1 + \lambda_2 \, {\cal U}_2
\otimes {\cal U}_2 + \lambda_3 \, {\cal U}_3 \otimes {\cal U}_3 =
\frac{1}{2} (J - \ci *J) \, , \\[2mm]
J \equiv - [\lambda_1 \, \Pi_1
\wedge \Pi_1 + \lambda_2 \, \Pi_2 \wedge \Pi_2 + \lambda_3 \, \Pi_3
\wedge \Pi_3] \, .
\end{array}
\end{equation}
Moreover, from lemma \ref{lemma-eigenvalues} we have $\rho =  e^{\ci
\theta} \sqrt{t}$, where $\theta$ is an arbitrary phase that gives a
duality rotation in the expression (\ref{w-can-I-6}) of the Weyl
tensor. Then, taking into account lemma \ref{lemma-I-6} and
expression (\ref{Pi-I-6-a}) we have:
\begin{theorem} \label{theorem-I-6}
If $T$is a BR tensor such that $\Tr T^2 = 0$ and $9 t^3 \equiv \Tr
T^3 \not=0$, then the Weyl tensor can be obtained, up to duality
rotation, as
\begin{equation} \label{W(T)}
W = \cos \theta \, W_0 +  \sin \theta \, *W_0 , \quad W_0 =
\sqrt{t}\, J  , \quad J_{\alpha \beta \mu \nu} \equiv 2(H_{\alpha
\mu \beta \nu} - H_{\alpha \nu \beta \mu}),
\end{equation}
where $H$ is the concomitant of the BR tensor given in expression
{\rm (\ref{H})} of lemma \ref{lemma-I-6}.
\end{theorem}

\begin{acknowledgements}
This work has been partially supported by the Spanish Ministerio de
Educaci\'on y Ciencia, MEC-FEDER project FIS2006-06062.
\end{acknowledgements}

\appendix

\section{Notation}  \label{Ap-notation}

\subsection{Products and other formulas involving 2-tensors $A$ and
$B$} \label{ap-2tensors}
\begin{enumerate}
\item Composition as endomorphisms: $A \cdot B $,
$$ {(A \cdot B)^{\alpha}}_{\beta}= {A^{\alpha}}_{\mu}
{B^{\mu}}_{\beta}$$
\item In general, for arbitrary tensors, $\cdot$ will be used to indicate the
contraction of adjacent indexes on the tensorial product.
\item Square and trace as endomorphism
$$A^2 = A \cdot A, \qquad \tr A = {A^{\alpha}}_{\alpha}$$
\item Action on vectors $x,\, y$ as an endomorphism $A(x)$ and as a bilinear form $A(x,y)$:
$$A(x)^{\alpha} ={A^{\alpha}}_{\beta} x^{\beta}, \qquad A(x,y) =
A_{\alpha \beta} x^{\alpha} y^{\beta} $$
\item Exterior product as double 1-forms: $A \wedge B$,
$$( A \wedge B)_{\alpha \beta \mu \nu} =
A_{\alpha \mu} B_{\beta \nu} + B_{\alpha \mu} A_{\beta \nu} -
A_{\alpha \nu} B_{\beta \mu} -B_{\alpha \nu} A_{\beta \mu}$$

\end{enumerate}

\subsection{Products and other formulas involving SD-endomorphisms
${\cal X}$ and ${\cal Y}$} \label{ap-SD}
\begin{enumerate}
\item Every self-dual (SD) symmetric double 2-form ${\cal X}$ defines a linear map on
the 3-dimensional SD bivector space. In short, we will say that
${\cal X}$ is a SD-endomorphism.
\item Composition as endomorphisms ${\cal X} \circ {\cal Y}$:
$$({\cal X} \circ {\cal Y})_{\alpha \beta \rho \sigma} = \frac{1}{2} \,
{{\cal X}^{\alpha \beta}}_{\mu \nu} {{\cal Y}^{\mu \nu}}_{\rho
\sigma}$$
\item{Square and trace as endomorphism}:
$$ {\cal X}^2= {\cal X} \circ {\cal X}, \qquad \Tr{\cal X} = \frac{1}{2}
{\cal X}^{\alpha \beta}_{\ \ \ \alpha \beta} $$
\item Action (on SD bivectors ${\cal F}$, ${\cal
H}$) as an endomorphism ${\cal X}({\cal F})$, and as a bilinear form
${\cal X}({\cal F},{\cal H})$:
$${\cal X}({\cal F})_{\alpha \beta} = \frac{1}{2} {{\cal X}_{\alpha \beta}}^{\mu \nu} \,
{\cal F}_{\mu \nu}, \qquad {\cal X}({\cal F},{\cal H})= \frac{1}{4}
{\cal X}_{\alpha \beta \mu \nu} {\cal F}^{\alpha \beta} {\cal
H}^{\mu \nu}$$
\item Metric on the space of SD bivectors (SD-identity):
$${\cal G}= \frac{1}{2} ( G - \ci \eta) \, , \qquad {\cal G}({\cal F}) = {\cal F}$$
\item The metric volume element $\eta$ is a linear map on the
2-forms space that defines the Hodge dual operator. For a real
2-from $F$ and a real symmetric double 2-form $W$:
$$*F = \eta(F) \, , \qquad *W = \eta \circ W \, .$$
\end{enumerate}

\subsection{Products and other formulas involving TLS-endomorphisms
$T$ and $R$.} \label{ap-TLS}
\begin{enumerate}
\item Every 4-tensor $T$ with the symmetries:
\begin{equation} \label{TLS-def}
T_{\alpha \beta \mu \nu} = T_{\beta \alpha \mu \nu} = T_{\mu \nu
\alpha \beta} \, , \qquad  T^{\alpha}_{\ \alpha \mu \nu} = 0
\end{equation}
defines a symmetric linear map on the 9-dimensional space of the
traceless symmetric tensors. We say that $T$ is a TLS-endomorphism.
\item Composition as endomorphisms: $T  \bullet  R$,
$${(T \bullet R )^{\alpha \beta}}_{\rho \sigma} = {T^{\alpha \beta}}_{\mu \nu}
{R^{\mu \nu}}_{\rho \sigma}$$
\item{Square and trace as endomorphism}:
$$T^2 = T \bullet T \, , \qquad  \Tr T = T^{\alpha \beta}_{\ \ \ \alpha \beta} $$
\item Action (on traceless symmetric tensors $A$, $B$) as an endomorphism $T(A)$
and as a bilinear form  $T(A, B)$,
$$T(A)_{\alpha \beta} = {T_{\alpha \beta}}^{\mu \nu} \, A_{\mu \nu}, \qquad T(A,B) =
T_{\alpha \beta \mu \nu} \, A^{\alpha \beta} B^{\mu \nu} $$
\item Metric on the space of traceless symmetric tensors (TLS-identity): $\Gamma$
$$\Gamma_{\alpha \beta \mu \nu} =\frac{1}{2} ( g_{\alpha \mu}  g_{\beta \nu} +
g_{\alpha \nu} g_{\beta \mu } ) - \frac{1}{4} g_{\alpha \beta}
g_{\mu \nu} \, , \qquad  \Gamma(A) = A$$
\end{enumerate}

\section{Frames} \label{Ap-frames}

\subsection{Frames of vectors} \label{Ap-f-vectors}
\begin{enumerate}
\item
In a spacetime with signature $(-,+,+,+)$  we shall note $\{
e_{\alpha} \} $ an orthonormal frame of vector fields having the
orientation $\eta = e_0 \wedge e_1 \wedge e_2 \wedge e_3 $.
\item
A (complex) null frame $\{ l, k , m, \overline{m} \}$ is defined
from the orthonormal frame by
\begin{equation}
l = \frac{1}{\sqrt{2}} (e_0 - e_1 ), \quad k = \frac{1}{\sqrt{2}}
(e_0 + e_1 ), \quad  m =\frac{1}{\sqrt{2}} (e_2 + \ci  e_3 )
\end{equation}
\end{enumerate}

\subsection{Frames of SD bivectors}  \label{Ap-f-bivectors}
\begin{enumerate}
\item
We shall note $\{ {\cal U}_i \}$  the orthonormal frame of SD
bivectors defined by
 \begin{equation}\label{u2f}
 {\cal U}_i = \frac{1}{\sqrt{2}} (U_i - \ci * U_i), \qquad U_i=e_0
 \wedge e_i
 \end{equation}
\item
The frame $\{ {\cal U}_i \}$ satisfies
 \begin{equation} \label{orient}
 {\cal U}_i \cdot {\cal U}_j = \frac{1}{2} \delta_{ij} \, g -
 \frac{\ci}{\sqrt{2}} \, \epsilon_{ijk} \, {\cal U}_k  , \quad
\epsilon_{123}=1
\end{equation}
\item The metric ${\cal G} = \frac{1}{2} (G - \ci \eta)$, in
the space of SD bivectors is ${\cal G} = - \sum {\cal U}_i \otimes
{\cal U}_i $.
\item
Associated with a null frame $\{ l, k, m, \overline{m} \}$ we have
the SD null frame $\{ {\cal U} , {\cal L}, {\cal K}
 \}$, defined by the relations
 \begin{equation}
{\cal U}= \frac{1}{\sqrt{2}} ( l  \wedge k - m \wedge \overline{m} )
, \quad {\cal L} =   l \wedge m , \quad {\cal K} = k \wedge
\overline{m}
\end{equation}
\item The metric ${\cal G}$ is given by ${\cal G} = - {\cal U} \otimes
{\cal U} - {\cal L} \stackrel{\sim}{\otimes} {\cal K} $.
\item
The SD frames $\{ {\cal U}_i \}$ and $\{ {\cal U} , {\cal L}, {\cal
K}  \}$ are related by
\begin{equation}
{\cal U} = {\cal U}_1, \quad {\cal L}= \frac{1}{\sqrt{2}} ({\cal
U}_2 + \ci {\cal U}_3)   , \quad {\cal K}= \frac{1}{\sqrt{2}} ({\cal
U}_2 - \ci {\cal U}_3)
\end{equation}
\end{enumerate}

\subsection{Frames of symmetric tensors}  \label{Ap-f-tensors}
\begin{enumerate}
\item
From every orthonormal frame $\{ {\cal U}_i \}$ of SD bivectors we
can define the (complex) orthogonal frame of traceless symmetric
tensors $\{ \Pi_{ij} \}$ defined as
\begin{equation}
\Pi_{ij} = {\cal U}_i \cdot \overline{\cal U}_j
\end{equation}
\item
The frame $\{ \Pi_{ij} \}$ satisfies
\begin{equation} \label{orient-tensors}
(\Pi_{ij}, \Pi_{km}) = \delta_{ik} \, \delta_{jm}
\end{equation}
\item
In terms of the frame $\{ e_{\alpha} \}$ the frame $\{ \Pi_{ij} \}$
takes the expression:
\begin{equation}
\Pi_i \equiv \Pi_{ii} = \frac{1}{2} ( v_i - h_i ), \qquad \Pi_{ij}=
\frac{1}{2} ( e_{i} \stackrel{\sim}{\otimes} e_j + \ci
\epsilon_{ijk}\,  e_{0} \stackrel{\sim}{\otimes} e_k )  \quad (i
\neq j)
\end{equation}
where $v_i = - e_0 \otimes e_0 + e_i \otimes e_i $, and $h_i = g - v_i$.
\item
The tern  $\{ \Pi_{i} \}$ satisfies
\begin{equation} \label{orient-tensors}
\Pi_{i}^2 = \frac14 \, g \, ; \qquad \Pi_i \cdot \Pi_j = \frac12 \,
\Pi_k \, , \quad (i,j,k \not=) \, .
\end{equation}
\item The metric $\Gamma$ in the space of traceless symmetric tensors is
$\Gamma = \sum_{i,j} \Pi_{ij} \otimes \Pi_{ij}$.
\item
From every null frame $\{ {\cal U} , {\cal L}, {\cal K} \}$ of SD
bivectors we can define the (complex) null frame of traceless
symmetric tensors $\{ \Pi, \Lambda, {\rm K}, {\rm N}, \overline{{\rm
N}}, \Omega, \overline{\Omega} , {\rm M}, \overline{{\rm M}} \}$
defined as
\begin{equation} \label{stnf}
\begin{array}{lll}
\Pi ={\cal U} \cdot \overline{\cal U}  \qquad &   \Lambda = {\cal L}
\cdot \overline{\cal L}  \qquad & {\rm K} = {\cal K} \cdot \overline{\cal K} \\
  {\rm N}={\cal U} \cdot \overline{\cal L}  \qquad &\Omega=  {\cal U}
\cdot
\overline{\cal K}   &{\rm M} =  {\cal L} \cdot \overline{\cal K} \\
\end{array}
\end{equation}
\item
In terms of the frame $\{ l, k,  m , \overline{m} \}$ the frame $\{
\Pi, \Lambda, {\rm K}, {\rm N}, \overline{{\rm N}}, \Omega,
\overline{\Omega} , {\rm M}, \overline{{\rm M}} \}$ takes the
expression:
\begin{equation} \label{nfsta}
\begin{array}{lll}
\Pi=- \frac{1}{2} ( l \stackrel{\sim}{\otimes}  k + m
\stackrel{\sim}{\otimes} \overline{m}), \qquad  &  {\rm N}= -
\frac{1}{\sqrt{2}} \, l \stackrel{\sim}{\otimes}\overline{m}, \qquad
&
\Omega= \frac{1}{\sqrt{2}} \, k \stackrel{\sim}{\otimes} m , \\[1mm]
{\rm M}  = m \otimes m ,\qquad &  \Lambda= - \, l \otimes l , \qquad
& {\rm K}= - \, k \otimes k
\end{array}
\end{equation}
\item The metric $\Gamma$ in the space of traceless symmetric tensors is
\begin{equation}
\Gamma= \Pi \otimes \Pi + \Lambda \stackrel{\sim}{\otimes} {\rm K} +
{\rm M} \stackrel{\sim}{\otimes}\overline{{\rm M}} + {\rm N}
\stackrel{\sim}{\otimes} \Omega + \overline{{\rm N}}
\stackrel{\sim}{\otimes} \overline{\Omega}
\end{equation}
\end{enumerate}

\section{Some technical results} \label{Ap-technical}

C.1.- Let $\{e_2, e_3\}$ be two orthonormal vectors of an
Euclidean vectorial space and let us suppose that the tensors $P= e_2 \otimes e_2 + e_3
\otimes e_3$ and $Q = e_2 \stackrel{\sim}{\otimes} e_3 $ are known.\\[1.5mm]
Firstly, we want to obtain  $\{e_2, e_3\}$ in terms of $Q$ and $V$. Let us define
$$\kappa = Q(x,x), \qquad \nu = V(x,x)$$
where $x$ is whatever vector such that $\kappa^2 - \nu^2 \neq 0$.
Let us take $\lambda_1$ one of the roots of the quadratic equation
$$\lambda^2 - \kappa \lambda + \frac14 \nu^2 =0$$
and let us define $p = \sqrt{\lambda_1}$, $q=\frac{\nu}{p}$. Then
$e_2$ and $e_3$ are given by
\begin{equation}  \label{Ac1}
e_2 = \frac{1}{p^2 - q^2} [ p P(x) - q Q(x) ] ,  \qquad
e_3 = \frac{1}{p^2 - q^2} [ p Q(x) - q P(x) ]
\end{equation}
Note that the pair $\{e_2, e_3\}$ is determined up to ordination and
sign, accordingly with the choice of the root of the quadratic
equation and the sign in the square root of $\lambda_1$.

From expressions (\ref{Ac1}), a straightforward calculation leads to:
\begin{lemma} \label{lemma-Ac1}
Let $\{e_2, e_3\}$ be two orthonormal vectors of an
Euclidean vectorial space. If $P= e_2 \otimes e_2 + e_3
\otimes e_3$ and $Q = e_2 \stackrel{\sim}{\otimes} e_3$, then:
$$
e_2 \otimes e_2 - e_3 \otimes e_3 = \frac{P(x) \otimes P(x) - Q(x)
\otimes Q(x)}{\sqrt{[P(x,x)]^2 - [Q(x,x)]^2}}
$$
where $x$ is an arbitrary vector such that $[P(x,x)]^2 \not= [Q(x,x)]^2$.
\end{lemma}
\ \\
C.2.- Let $\{e_1, e_2, e_3\}$ be three orthonormal vectors of an
Euclidean vectorial space and let us suppose that one knows the tensors
\begin{equation} \label{A-B-C}
\displaystyle A= \sum_{i=1}^3 e_i \otimes e_i , \quad B=\sum_{\sigma
\in \Sigma_3} e_{\sigma(1)} \otimes e_{\sigma(2)} \otimes
e_{\sigma(3)} , \quad C= \sum_{i<j}  [ e_i \stackrel{\sim}{\otimes}
e_j ] \otimes [ e_i \stackrel{\sim}{\otimes} e_j] .
\end{equation}
Let us define
\begin{equation} \label{a-b-c}
\alpha=\frac13 A(x,x),  \quad \beta = \frac{1}{6} B(x,x,x) \ \quad \gamma
= \frac{1}{12} C(x,x,x,x)
\end{equation}
where $x$ is whatever vector such $\Delta \not= 0$ where
\begin{equation} \label{delta-p-q}
\Delta \equiv \sqrt{r^2+s^3} \, , \quad r \equiv \gamma-\alpha^2 \,
, \quad s \equiv \frac12(3 \alpha \gamma- \beta^2)- \alpha^3 \, .
\end{equation}
Then, the tensors
\begin{equation} \label{E-I}
E = A(x) \otimes A(x) - \frac{1}{2} C(x,x) \, , \qquad  I =
\frac{1}{4} B(x,x) \otimes B(x,x) - \beta \,  B(x) \, ,
\end{equation}
take, in terms of $e_i$, the expressions:
\begin{equation} \label{E-I-es}
E = \sum_{i=1}^3 y_i \, e_i \otimes e_i \, , \qquad  I = \sum_{(ijk
)} y_i y_j \, e_{k} \otimes e_{k} \, ,
\end{equation}
for $(i,j,k)$ a pair permutation of $(1,2,3)$, and where $y_i$ are
the solutions of the cubic equation
$$y^3 - 3 \alpha y^2  + 3 \gamma y - \beta^2 =0$$

Then, if one writes the solutions $y_i$ in terms of the cubic roots
of the unity, a straightforward calculation leads to:
\begin{lemma} \label{lemma-Ac2}
Let $\{e_1, e_2, e_3\}$ be three orthonormal vectors of an Euclidean
vectorial space. Let $A$, $B$ and $C$ be the tensors given in {\rm
(\ref{A-B-C})} and let $\lambda_i$ be the cubic roots of the unity.
Then:
$$
\sum_{i=1}^3 \lambda_i \, e_i \otimes e_i
 = \frac{1}{2w \Delta}\left[-(\alpha w^3 +2
r\alpha^2+ r^2)A +(w^3+r\alpha)E+ r I \right]
$$
where the tensors $E$ and $I$ are given in {\rm (\ref{E-I})} and the
scalars $\alpha$, $\beta$ and $\gamma$ are defined in {\rm
(\ref{a-b-c})}, $x$ being an arbitrary vector such that $\Delta
\not= 0$, and where  $r$, $s$ and $\Delta$ are defined in {\rm
(\ref{delta-p-q})}, and $w$ is a nonvanishing scalar defined by one
of the expressions $w^3 \equiv - s \pm \Delta$.
\end{lemma}
Note that the expression obtained in lemma above is determined up to
ordination of the three $\lambda_i$, accordingly with the choice of
the root in expression of $w^3$. In this expression the sign before
$\Delta$ must be chosen such that $w \not=0$.


\begin{thebibliography}{99}



\bibitem{bel-1} Bel, L.: C. R. Acad. Sci. {\bf 247}, 1094 (1958)

\bibitem{bel-2} Bel, L.: C. R. Acad. Sci. {\bf 248}, 1297 (1959)

\bibitem{bel-3} Bel, L.: Cah. de Phys. {\bf 16}, 59 (1962). This article
has been reprinted in Gen. Rel. Grav. {\bf 32}, 2047 (2000)

\bibitem{seno} Senovilla, J.M.M.: Class. Quantum Grav. {\bf 17}, 2799 (2000)

\bibitem{garcia} Garc\'{\i}a-Parrado, A.: Class. Quantum Grav. {\bf 25}, 015006 (2008)

\bibitem{debever-1} Debever, R.: Bulletin de la Societ\'e Math\'ematique de Belgique
{\bf t. X}, 112 (1958)

\bibitem{debever-2} Debever, R.: C. R. Acad. Sci. {\bf 249}, 1324 (1959)

\bibitem{bergqvist-lan-1} Bergqvist, G., Lankinen, P.: Class. Quantum Grav. {\bf 21},
3499 (2004)

\bibitem{rai} Rainich, G.Y.: Trans. Am. Math. Soc. {\bf 27}, 106
(1925)

\bibitem{fsY} Ferrando, J.J., S\'aez, J.A.: Gen. Relativ. Gravit. {\bf 35},
1191 (2003)

\bibitem{fsBR1} Ferrando, J.J., S\'aez, J.A.: Gen. Relativ. Gravit. {\bf 41},
(published on line) (2009)

\bibitem{kra} Stephani, H., Kramer, D., MacCallum, M., Hoenselaers, C., Herlt, E.:
{\it Exact solutions to Einstein's field equations} (Cambridge
University Press, Cambridge, 2003)

\bibitem{petrov-W} Petrov, A.Z.: Sci. Not. Kazan Univ. {\bf 114}, 55 (1954). This article
has been reprinted in Gen. Rel. Grav. {\bf 32}, 1665 (2000)

\bibitem{mcar} McIntosh, C.B.G., Arianrhod, R.: Class. Quantum Grav.
{\bf 7}, L213 (1990)

\bibitem{fsI} Ferrando, J.J., S\'aez, J.A.: Class. Quantum Grav. {\bf 14}, 129 (1997)

\bibitem{fs-aligned-em} Ferrando, J.J., S\'aez, J.A.: Gen. Relativ. Gravit. {\bf 36},
2497 (2004)

\bibitem{fsWem} Ferrando, J.J., S\'aez, J.A.: Class. Quantum Grav. {\bf 19}, 2437 (2002)

\bibitem{fms} Ferrando, J.J., Morales, J.A., S\'aez,
J.A.: Class. Quantum Grav. {\bf 18}, 4969 (2001)



\end{thebibliography}
\end{document}